\documentclass[aps,prl,showpacs,twocolumn,floatfix]{revtex4}
\input epsf.sty
\def\MP{\mbox{$M_P$}}

\def\mbh{\mbox{$M_{\rm BH}$}\ }     % M_BH with the space at the end
\def\MBH{\mbox{$M_{\rm BH}$}}         % no space at the end
\def\MET{\mbox{${\hbox{$E$\kern-0.6em\lower-.1ex\hbox{/}}}_T$}} %missing ET
\def\met{\mbox{${\hbox{$E$\kern-0.6em\lower-.1ex\hbox{/}}}_T$}\ } %missing ET w/ space at the end
\newcommand{\AmS}{{\protect\the\textfont2
   A\kern-.1667em\lower.5ex\hbox{M}\kern-.125emS}}

\begin{document}
\title{Discovering New Physics in the Decays of Black Holes}
\vspace*{-0.2in}

\author{Greg Landsberg}
\affiliation{Department of Physics, Brown University, Providence, RI 02912, USA}

\begin{abstract}
If the scale of quantum gravity is near a TeV, the LHC will be
producing one black hole (BH) about every second, thus qualifying
as a BH factory. With the Hawking temperature of a few 
hundred GeV, these rapidly evaporating BHs may produce new, 
undiscovered particles with masses $\sim 100$ GeV. The probability 
of producing a heavy particle in the decay depends on its mass only weakly,
in contrast with the exponentially suppressed direct production.
Furthemore, BH decays with at least one prompt charged lepton 
or photon correspond to the final states with low background. Using 
the Higgs boson as an example, we show that it may be found at the 
LHC on the first day of its operation, even with incomplete detectors.
\end{abstract}

\maketitle

{\bf Introduction:} An exciting consequence of TeV-scale quantum
gravity \cite{add} is the possibility of production of black holes
(BHs)~\cite{adm,bf,ehm} at CERN's Large Hadron Collider (LHC) 
and beyond. As was shown recently~\cite{dl,gt}, the cross section 
fo BH production at the LHC is $\sim$100~pb for a
fundamental Planck scale ($M_P$) $\sim$1~TeV, which would turn 
the LHC into a BH factory with a production rate $\sim$1~Hz.

Once produced, the TeV BHs quickly evaporate via Hawking 
radiation~\cite{Hawking} into half-a-dozen of particles~\cite{dl,gt}. 
Since gravity couples ``democratically" to different particle species, 
BHs would evaporate predominantly into particles on our 
brane~\cite{ehm}, with relative probabilities of emitting 
different particles depending mainly on particle quantum numbers
(e.g., spin and color), and only slightly on particle mass,
as long as it is below the Hawking temperature. For a typical BH with 
the mass $\sim$1~TeV, the Hawking temperature is a few 
hundred GeV~\cite{dl}, which opens the possibility of producing new
particles with masses $\sim 100$~GeV in decays of BHs.

This has exciting consequences for searches for new physics at the 
LHC and beyond, as the production cross section for any new particle via
this mechanism is (i) large, and (ii) depends only weakly on particle mass, 
in contrast with the exponentially suppressed direct 
production mechanism. In this Letter, we consider a search for an 
intermediate-mass Higgs boson, most challenging to observe at the 
Tevatron~\cite{SUSY-Higgs}, as an exciting example of this possibility. 
We show that for $M_P \sim 1$~TeV, a 130 GeV Higgs boson can be observed 
in decays of BHs at the LHC in as little as one hour of operation, even 
with incomplete detectors.

{\bf Black Hole Production and Decay:} 
The parton-level black hole production cross section is given by~\cite{dl}:
\vspace*{-0.1in}
$$
    \sigma(\MBH) \approx \pi R_S^2 = \frac{1}{M_P^2}
    \left[
      \frac{\MBH}{\MP} 
      \left( 
        \frac{8\Gamma\left(\frac{n+3}{2}\right)}{n+2}
      \right)
    \right]^\frac{2}{n+1},
$$
where $n$ is the number of large extra dimensions, $R_S$ is the Schwarzschild 
radius of the BH, and \mbh is its mass. In order to obtain the 
production cross section in $pp$ collisions at the LHC, we use the parton 
luminosity approach~\cite{EHLQ,dl,gt}:
\begin{equation}
    \frac{d\sigma(pp \to \mbox{BH} + X)}{d\MBH} = 
    \frac{dL}{dM_{\rm BH}} \hat{\sigma}(ab \to \mbox{BH})
    \left|_{\hat{s}=M^2_{\rm BH}}\right.,\label{cs}
\end{equation}
where the parton luminosity $dL/d\MBH$ is defined as the sum over
all the types of initial partons:
$$
    \frac{dL}{dM_{\rm BH}} = \frac{2\MBH}{s} 
    \sum_{a,b} \int_{M^2_{\rm BH}/s}^1  
    \frac{dx_a}{x_a} f_a(x_a) f_b(\frac{M^2_{\rm BH}}{s x_a}),
    \vspace*{-0.1in}
$$
and $f_i(x_i)$ are the parton distribution functions (PDFs). We
used the MRSD$-'$~\cite{MRSD} PDFs with the $Q^2$ scale taken
to be equal to \MBH, which is within the allowed range of these
PDFs for up to the kinematic limit at the LC. The dependence of 
the cross section on the choice of PDF is $\sim 10\%$.

As we expect unknown quantum gravity effects to play an increasingly 
important role for the BH mass approaching the fundamental Planck scale,
following the prescription of Ref.~\cite{dl}, we do not consider BH 
masses below the Planck scale. It is expected that the BH 
production rapidly turns on, once the relevant energy threshold
$\sim\! M_P$ is crossed. (At lower energies, we expect BH production
to be exponentially suppressed due to the string excitations or 
other quantum effects.)  

The total production cross section for $\MBH > M_P$ at the LHC, 
obtained from Eq.~(\ref{cs}), ranges between 15 nb and 1 pb for 
the Planck scale between 1 TeV and 5 TeV, and varies by $\approx 10\%$ 
for $n$ between 2 and 7.

Once produced, the BHs evaporate at the Hawking temperature $T_H$ given 
by~\cite{mp}:
\vspace*{-0.1in}
\begin{equation}
    T_H = \MP
    \left(
      \frac{\MP}{\MBH}\frac{n+2}{8\Gamma\left(\frac{n+3}{2}\right)}
    \right)^\frac{1}{n+1}\frac{n+1}{4\sqrt{\pi}}
\label{TH}
\end{equation}
As the parton collision energy increases, the resulting BH gets 
heavier and its decay products get colder. The average multiplicity 
of particles produced in the BH evaporation is given by~\cite{dl}: 
$\langle N \rangle \approx \frac{\MBH}{2T_H} \sim 5$. Note, that due 
to the rapidly falling PDFs, most of the BHs are produced with $\MBH 
\approx M_P$, and therefore the Hawking temperature given by 
Eq.~(\ref{TH}) increases roughly proportional to the Planck mass; 
consequently, the average decay multiplicity stays approximately 
constant.

The decay of a BH is thermal: it obeys all local conservation laws, but 
otherwise does not discriminate between particle species. Since the typical 
decay involves a large number of particles, the BH emits all the 
$\approx 60$ SM particles with roughly equal probability. Since there are 
six charged leptons and one photon, we expect $\sim 10\%$ of the decay products 
to be hard, primary charged leptons and $\sim 5\%$ of them to be 
hard photons or $W/Z$ bosons, each carrying hundreds of GeV of energy. Similarly, 
approximately 75\% of particles produced in BH decays are quarks 
and gluons (due to the color-factor enhancement), while $\sim 1\%$ of 
them are Higgs bosons, provided that they are sufficiently light.

A relatively large fraction of prompt and energetic photons, electrons, 
and muons are expected in the high-multiplicity BH decays, which would 
make it possible to select pure samples of BH events, which are also 
easy to trigger on~\cite{dl,gt}. These can be used to search for other new 
particles that could appear in the BH decays at rates similar 
to those for the SM species.

{\bf Higgs Boson Production in the Black-Hole Decays:} 
As an example, we use the production of a SM-like Higgs boson with 
the mass of 130 GeV, still allowed in low-scale supersymmetry models, but very 
hard to establish at the Fermilab Tevatron~\cite{SUSY-Higgs}. We consider decay 
of the Higgs boson into pair of jets, dominated by the $b\bar b$ final state 
(57\%), with an additional 10\% contribution from the $c\bar c$, $gg$, and 
hadronic $\tau\tau$ final states.

We model the production and decay of the BH with the TRUENOIR Monte Carlo 
generator~\cite{Snowmass}, which implements a euristic algorithm to describe a 
spontaneous decay of a BH. The generator is interfaced with the PYTHIA Monte 
Carlo program~\cite{PYTHIA} to account for the effects of initial and final state radiation, particle decay, and fragmentation. We used a 1\% probability 
to emit the Higgs particle in the BH decay, which is likely to be a 
conservative estimate, as the emission of a spin-0 particle is 
expected to be enhanced by the grey body factors~\cite{grey-body}. While the
multidimensional grey body factors are yet to be calculated, the 4-dimensional 
analysis shows that the emission of the Higgs boson can be enhanced by as much as
a factor of 2.5~\cite{lykken}.

We reconstruct electron, muons, photons, and jets within the rapidity range of 
$\pm 2.5$ and require the transverse energy above 50 GeV for leptons and photons, 
and 25 GeV for jets. Jets are reconstructed using a parton-level clustering algorithm and 
then smeared assuming energy resolution $\sigma(E)/E = 0.60/\sqrt{E/\mbox{GeV}} 
\oplus 0.02$~\cite{ATLAS}, typical of an LHC detector. Note, that the high boost of the particles produced in the BH decay with the average energy $\sim T_H$ improves the dijet mass resolution. For the studies that involve $b$-tagging, we assume the tagging
efficiency of 60\%, with a 5\% probability of mistakenly tagging a non-$b$ 
jet; this high mistagging probability reflects an enhanced charm quark content 
in the BH decays.

In order to select a pure sample of BH events, we used the prescription of
Ref.~\cite{dl} by selecting only the events with four or more objects 
(i.e., electrons, muons, photons, or jets) in the final state, with at 
least one of them being not a jet. This selection reduces SM backgrounds
to a negligible level, see Fig.~2 in Ref.~\cite{dl}, and results in a 
typical signal acceptance of 7\%, dominated by the probability to find 
an isolated prompt lepton or photon in the event. (A Higgs boson is 
found in about 5\% of these tagged events.)

\begin{figure}[tbp]
\begin{center}
\epsfxsize=3.3in
\epsffile{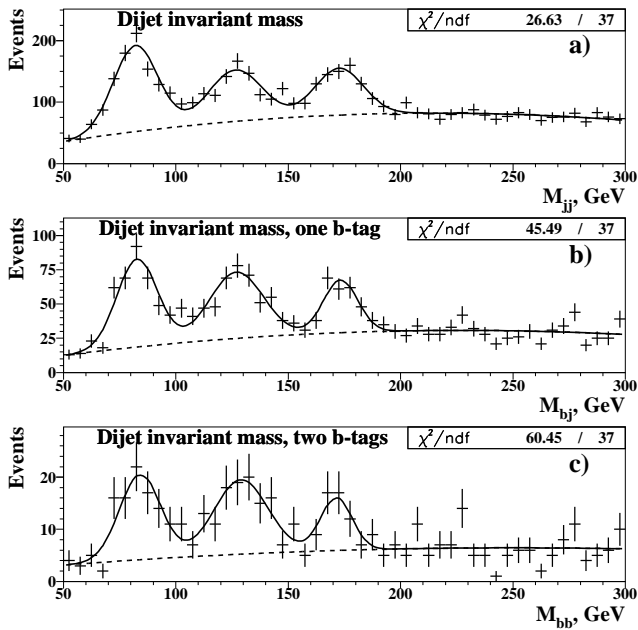}
\vspace*{-0.1in}
\caption{Dijet invariant mass observed in the BH decays with a prompt 
lepton or photon tag in $\approx$3~pb$^{-1}$ of the LHC data, for 
$M_P = 1$~TeV and $n=3$: (a) all jet combinations; (b) jet combinations 
with at least one of the jets tagged as a $b$-jet; (c) jet 
combinations with both jets tagged as $b$-jets. The solid line is a 
fit to a sum of three Gaussians and a polynomial background, shown 
with the dashed line. The three peaks correspond to the $W/Z$ bosons, 
the Higgs boson, and the top quark (see text). The $\chi^2$ per d.o.f.
is shown to demonstrate the quality of the fit. Note, that as the 
maximum likelihood fit was used for all cases, the $\chi^2$ in (c) is 
not an appropriate measure of the fit quality due to low statistics. 
Using the Poisson statistics, the probability of the fit (c) is 8\%.}
\vspace*{-0.35in}
\label{fig1}
\end{center}
\end{figure}

The dijet invariant mass spectrum in the decays of BHs with the masses 
above $M_P = 1$~TeV is shown in Fig.~\ref{fig1}. The three panes correspond 
to all jet combinations (with the average of approximately four jet 
combinations per event), combinations with at least one $b$-tagged jet, 
and combinations with both jets $b$-tagged. The most prominent feature 
in all three plots is the presence of three peaks with the masses around 
80, 130, and 175 GeV. The first peak is due to the hadronic decays of the 
$W$ and $Z$ bosons produced in the BH decay either directly or in the decays 
of the top and Higgs particles. (The resolution of a typical LHC detector 
does not allow to resolve $W$ and $Z$ in the dijet mode.) The second peak 
is due to the $h \to jj$ decays, and the third peak is due to the $t \to 
Wb \to jjb$ decays, where the top quark is highly boosted. In this case, 
one of the jets from the $W$ decay sometimes overlaps with the prompt 
$b$-jet from the top quark decay, and thus the two are reconstructed as 
a single jet; when combined with the second jet from the $W$ decay, this 
gives a dijet invariant mass peak at the top quark mass. The data set
shown in Fig.~\ref{fig1} consists of 50K BH events, which, given the 15~nb 
production cross section, corresponds to an integrated luminosity of 
3 pb$^{-1}$, or less than an hour of the LHC operation at the nominal 
luminosity.

We calculate the signal significance via a counting experiment in the 
optimal window of $\pm 2\sigma$ around the signal Gaussian~\cite{bosonic}. 
Given the signal $S$ and the background $B$ in this window, we define 
the significance as $S/\sqrt{B+(\delta B)^2}$, where $\delta B$ is the 
uncertainty on the background, as determined from the fit. The 
significance of the Higgs signal shown in Fig.~\ref{fig1}a is 6.7$\sigma$.

The effect of $b$-tagging is shown in the other two panes of 
Fig.~\ref{fig1}. Only the jet pair combinations with at least one 
of the two jets tagged are shown in Fig.~\ref{fig1}b; the pairs with 
both jets tagged are shown in Fig.~\ref{fig1}c. While $b$-tagging 
improves the signal-to-background ratio, the Higgs signal significance 
with a single or double $b$-tagging deteriorates slightly (6.6$\sigma$ 
and 5.4$\sigma$, respectively). The possibility to discover the Higgs 
boson in the $b\bar b$ decays without $b$-tagging has an important 
impact on the LHC collaborations, as the vertex detector installation
is likely to be staged. Moreover, even with the installed detectors, 
it takes a large amount of data and a long time to understand and 
align them, so that high $b$-tagging efficiency can be achieved. 
One would not expect this to happen before several months of stable 
running of the LHC. On the other hand, the discovery in the non-tagged 
channel can happen almost immediately after the LHC turn-on, using
preliminary calibrated calorimeters only.

\begin{figure}[tbp]
\begin{center}
\epsfxsize=3.3in
\epsffile{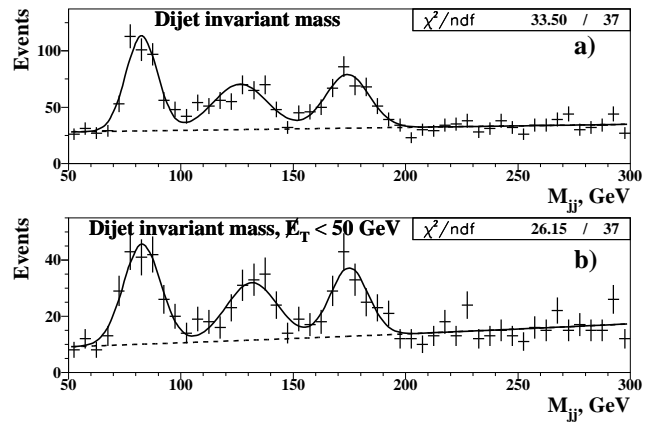}
\vspace*{-0.1in}
\caption{Dijet invariant mass observed in the BH decays with a prompt 
lepton or photon tag in $\approx$100~pb$^{-1}$ of the LHC data, for 
$M_P = 2$~TeV and $n=3$: (a) all jet combinations; (b) same, for a 
subset of the events with $\MET < 50$~GeV. See Fig.~\protect\ref{fig1}
caption for explanation.}
\vspace*{-0.35in}
\label{fig2}
\end{center}
\end{figure}

Another way to improve the Higgs boson identification is to select only 
the events with a low amount of missing energy, which guarantees that 
the jets are not grossly mismeasured. The effect of a $\MET < 50$~GeV 
cut is shown in Fig.~\ref{fig2}, based on 50K BH events for $M_P = 2$~TeV 
and $n=3$, which corresponds to $\sim$100~pb$^{-1}$ of data, given the 
452~pb production cross section. The top pane is the same as in 
Fig.~\ref{fig1}a, while the bottom pane shows a slight improvement in 
the signal-to-background ratio after the \met cut is applied.

\begin{figure}[tbp]
\begin{center}
\epsfxsize=3.3in
\epsffile{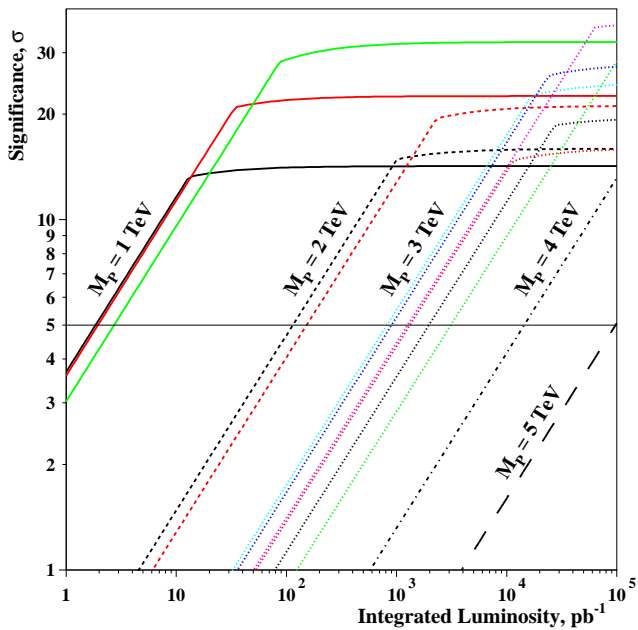}
\vspace*{-0.1in}
\caption{
The significance of a 130~GeV Higgs discovery in the black-hole decays at the LHC 
as a function of the integrated luminosity. The solid curves correspond 
to $M_P = 1$~TeV, $n=3$ and (left to right) no $b$-tagging, a single $b$-tag, 
two $b$-tags. The two short-dashed lines correspond to $M_P = 2$~TeV, $n=3$, no $b$-tagging, and (left to right) no \met cut, $\MET < 50$~GeV cut. The six dotted lines correspond to $M_P = 3$~TeV, no $b$-tagging, and (left to right) $n = 7$, 5, 6, 2, 3, and 4. The dashed-dotted and long-dashed lines correspond to $M_P = 4$ and 5 TeV, respectively, $n=3$, and no $b$-tagging. The solid horizontal line indicates the $5\sigma$ discovery
level.}
\vspace*{-0.35in}
\label{fig3}
\end{center}
\end{figure}

In order to find Higgs discovery significance as a function of the 
integrated luminosity ($\cal L$), we use the above significance definition. 
Both the signal and background are scaled with the integrated luminosity. 
Since the background can be determined from the dijet mass distribution 
away from the signal region, we assume that the error on the background is 
dominated by the statistics and scales as $1/\sqrt{\cal L}$, until it 
becomes limited by the systematics due to the shape of the fitting function, 
which is taken to be 5\%. Constant fractional systematic uncertainty 
results in a plateau in signal significance, as its luminosity dependence
cancels out when the systematics dominates. The results are shown in 
Fig.~\ref{fig3} for several choices of $M_P$, between 1 and 5 TeV.

The effect of $b$-tagging is shown in Fig.~\ref{fig3} with the family 
of curves that correspond to $M_P = 1$~TeV. While the data without 
$b$-tagging (the leftmost curve) gives a slightly higher significance, 
it reaches the plateau at lower values of the significance than the 
single $b$-tag (middle curve) and double $b$-tag (the rightmost curve) 
data. This is due to a higher signal-to-background ratio in the tagged 
sample. The effect of a $\MET < 50$~GeV cut is illustrated with the two curves 
that correspond to $M_P = 2$~TeV. Again, while the significance is 
slightly higher for the full sample, the saturation is achieved at higher 
values for the sample with the $\MET$ cut.

Finally, we study the effect of the number of extra dimensions for 
$M_P = 3$~TeV. This dependence is expected to be complicated, 
as the cross section, multiplicity of the final state particles, and 
Hawking temperature all depend on $n$~\cite{dl}. The six curves for 
$M_P = 3$~TeV correspond (left to right) to $n = 7$, 5, 6, 2, 3, and 4. 
(As expected, $n=4$ corresponds to the lowest, and $n=7$ to the highest 
cross section.) The difference between the best and the worst cases 
corresponds to approximately a factor of two in the integrated luminosity 
required to achieve a given significance level. The $n=3$ case mostly used 
in this Letter is on the conservative side.

The $5\sigma$ discovery of a 130 GeV Higgs boson may be possible with 
${\cal L} \approx 2$~pb$^{-1}$, 100~pb$^{-1}$, 1~fb$^{-1}$, 10~fb$^{-1}$, 
and 100~fb$^{-1}$ for the fundamental Planck scale of 1, 2, 3, 4, and 
5~TeV, respectively. The amount of data required is significantly lower 
than that in a direct Higgs boson production, if the Planck scale is 
below 4 TeV. While the studies were done for a particular Higgs boson 
mass, the dependence on the mass is small, and as seen from 
Figs.~\ref{fig1},~\ref{fig2}, a large range of masses for which the 
decay of the Higgs boson into $b\bar b$ dominates, can be probed. 
For the Higgs boson masses above 140 GeV, the $ZZ^*$ decay mode can be 
employed, similar to the approach used in the direct searches. (The
$ZZ^*$ mode would result in a significantly reduced combinatorial and 
in a negligible top quark backgrounds.)

Moreover, the approach outlined in this paper is applicable to searches 
for other new particles with the masses $\sim 100$~GeV, particularly 
the low-scale supersymmetry. Light slepton or top squark searches via 
this technique may be particularly fruitful, especially since the 
probability of emitting a spin-0 particle in the BH decays is expected 
to be enhanced. Finally, while we considered only the BH production 
and decay, very similar conclusions apply to the intermediate quantum 
states, such as string balls~\cite{de}, which have similar production 
cross section and decay modes as BHs. In this case, the relevant mass
scale is not the Planck scale, but the string scale, which determines 
the evaporation tempearature~\cite{de}.

{\bf Summary:} Black hole production at the LHC may change the way we
search for new particles. Decays of heavy BHs, tagged with
prompt leptons or photons, offer low-background environment for searches 
of new particles with mass $\sim 100$~GeV. For example, a 130~GeV 
SM-like Higgs boson may be observed with the significance of five standard 
deviations in one hour, day, month, or year of the LHC operation for 
the values of the fundamental Planck scale of 1, 2, 3, and 4 TeV, 
respectively. Thus, future high-energy colliders may become not only the 
black-hole factories, but factories for new physics.

{\bf Acknowledgments:} I would like to thank Savas Dimopoulos, 
Roberto Emparan, and Nemanja Kaloper for valuable discussions.
Special thanks to Steve Mrenna for help with the interfacing of 
the TRUENOIR code with PYTHIA, as well as to the Snowmass 2001
organizers, where these studies have been started. This work 
was supported partially by the U.S.~Department of Energy under 
Grant No. DE-FG02-91ER40688 and by the Alfred P. Sloan Foundation.

\end{document}